\documentclass[aip,apl,reprint,groupedaddress]{revtex4-1}

\usepackage{graphicx,subfigure}
\usepackage[percent]{overpic}
\graphicspath{{Figures/}}
\begin{document}


\title{A non-invasive electron thermometer based on charge sensing of a quantum dot}



\author{A. Mavalankar}
\email[]{am877@cam.ac.uk}
\author{S.J. Chorley}
\author{J. Griffiths}
\author{G.A.C. Jones}
\author{I. Farrer}
\author{D.A. Ritchie}
\author{C.G. Smith}
\affiliation{Semiconductor Physics Group, Department of Physics, University of Cambridge}


\date{\today}

\begin{abstract}
We present a thermometry scheme to extract the temperature of a 2DEG by monitoring the charge occupation of a weakly tunnel-coupled `thermometer' quantum dot using a quantum point contact detector. Electronic temperatures between 97 mK and 307 mK are measured by this method with an accuracy of up to 3 mK, and agree with those obtained by measuring transport through a quantum dot. The thermometer does not pass a current through the 2DEG, and can be incorporated as an add-on to measure the  temperature simultaneously with another operating device. Further, the tuning is independent of temperature.
\end{abstract}

\pacs{81.05.Ea,85.35.Gv,07.20.Dt,73.23.Hk,73.63.Kv}

\maketitle

The two-dimensional electron gas in GaAs/AlGaAs hetero-structures has diverse applications at cryogenic temperatures. For example, both gate-defined quantum dots and non-abelian fractional quantum hall states are candidates for a solid state quantum computer \cite{elzerman,fqhe}. Almost all these applications require a low operating temperature, which is achieved by cooling the device using a dilution refrigerator or a helium-3 system. These external cooling mechanisms rely on the lattice cooling the electron gas via an exchange of phonons. In the mK range and lower, phonon coupling decreases with temperature, and this reduces the ability of external cooling mechanisms to cool the 2-DEG, limiting the minimum achievable temperature \cite{price,ridley,roukes}. In practice, the electron gas is heated by unintended noise in the measurement set up to a temperature higher than the lattice. In many cases, the background electron temperature of an operating device is needed to analyze its behavior. This temperature has to be measured before or after the main experiment, an approach which cannot take into account any drift in electron temperature which occurs during the experiment. Our work describes the implementation of a thermometer which can perform an accurate and non-invasive temperature measurement \cite{prancethesis,theory}. The thermometer does not draw current from the sample being measured, which minimizes any back-action on the main experiment. 

Several low temperature thermometers have recently been realized using modern microlithography and nanolithography techniques \cite{review_therm}. For example, a fast NIS (normal metal-insulator-superconductor) junction thermometer, which operates at RF frequencies, has measured temperatures from 300 mK to 950 mK \cite{schmidt}. A `Shot Noise Thermometer' (SNT) which extracts the electron temperature from the electrical shot noise through a tunneling junction between two metal electrodes has been shown to operate over a temperature range of 30 mK to 300 K with an accuracy of 0.1 $\%$ in the middle two decades of that range \cite{spietz}. A `Coulomb Blockade Thermometer' (CBT) which relies on the temperature-dependent conductance of an array of tunneling junctions in an Al/Al$_2$O$_3$/Al structure, and which has an operating range of 20mK to tens of Kelvin with an absolute accuracy of 1$\%$ \cite{meschke,pekola}, has been developed into a commercial product. The CBT can also operate in a magnetic field, which is an important consideration for microthermometers used in low-temperature experiments.

Although these low temperature thermometers are an exciting area of research in themselves, they cannot measure the temperature of electrons in operating devices, and determining this electron temperature continues to be an active area of research \cite{rossi,italy,yacoby}. The electron temperature in these devices is usually extracted by measuring the current or conductance through a quantum dot when transport through it is Coulomb blockaded, and fitting to the resulting temperature dependent line shape \cite{beenakker,CB_lineshapes3}. In this paper, we demonstrate a non-invasive thermometer which does not pass a current through the area being measured. Additionally, it is easy to fabricate and operate, and special electronics like RF lines and cryogenic pre-amplifiers, or different materials like superconductors are not required.

\begin{figure*}
 \includegraphics{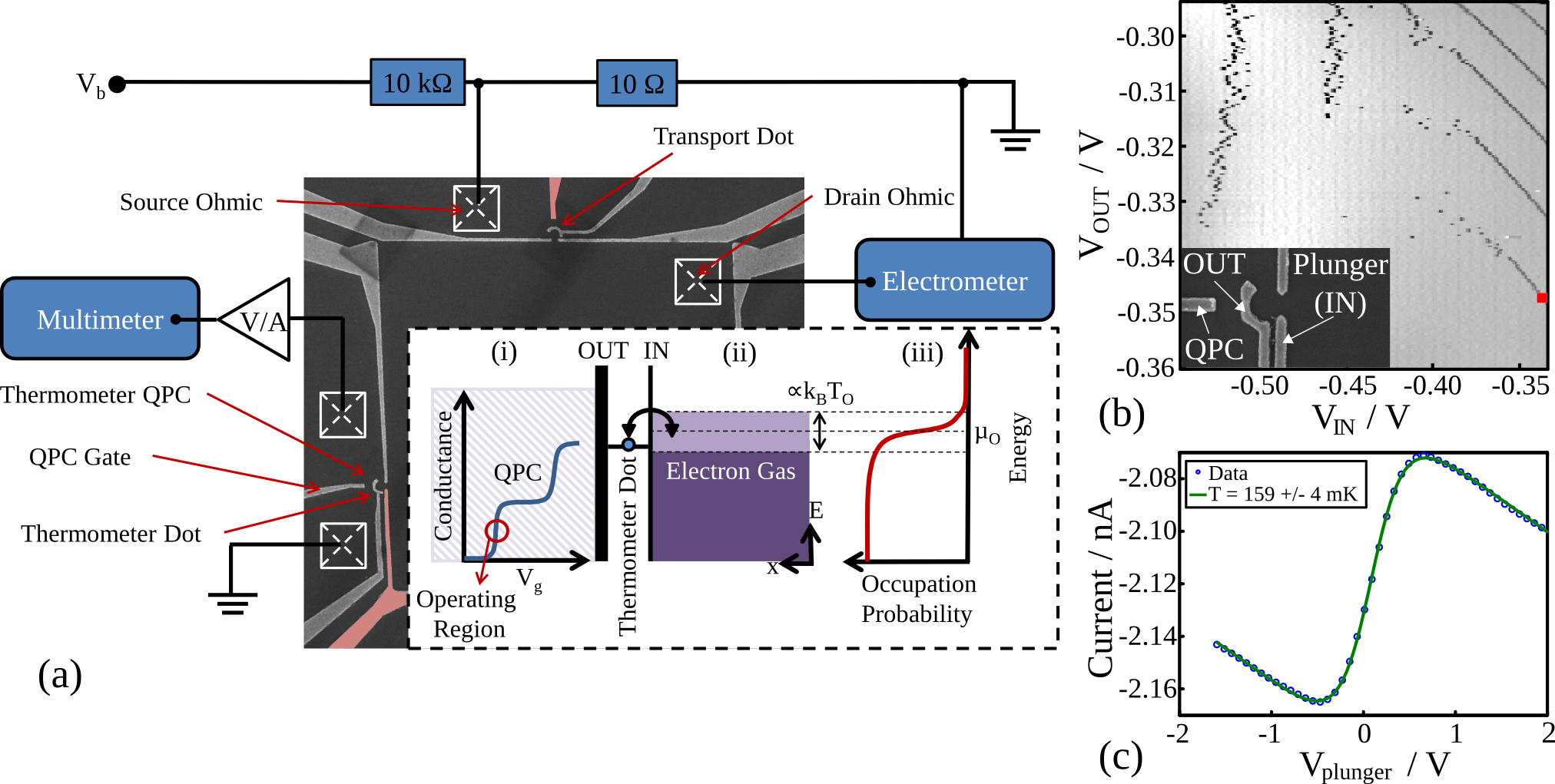}
 \caption{(a) Scanning Electron Micrograph of the device (similar to the one used in measurements). The circuit on the left monitors the charge distribution of the thermometer quantum dot. The circuit on the right simultaneously measures the current due to an applied bias through the Coulomb blockaded right dot. The gates colored red are used to vary the energy levels of the two dots through their respective bias windows. The inset is a schematic of the thermometry set-up. (i) The quantum point contact monitors the charge distribution of the thermometer dot. (ii) The dot is tuned so that it is opaque to the 2DEG on the left and partially transparent to the 2DEG on the right. (iii) The width of the resulting Fermi-Dirac signal gives the temperature of the electron gas. (b) A greyscale plot of the qpc transitions as a function of the voltages on the IN and OUT gates (see inset). The color scale is the current through the qpc differentiated with respect to $\mathrm{V_{IN}}$. The red square approximately marks the voltages at which the temperature measurements are made. (c) The IN gate is also used as a plunger gate to vary the energy level of the thermometer dot past the Fermi Dirac distribution of the electron gas. The blue circles are measured data (which is the average of the same measurement made 9 times), and the green line is the best fit calculated by optimizing the sum of least squares. The temperature obtained from this data set was 159$\mathrm{\pm}$4 mK.}
\label{fig:schematic}
\end{figure*}

Figure \ref{fig:schematic}(a) shows a scanning electron micrograph of the measured device, along with the electrical circuit used experiments. Ti/Au gates were patterned on the surface of a GaAs/AlGaAs heterostructure using electron-beam lithography. The 2DEG resided 90 nm below the surface, and was contacted by annealing AuGeNi ohmic contacts. The mobility and carrier concentration of the 2DEG are measured to be $\mathrm{1.25\cdot 10^{6} cm^{2} V^{-1} s^{-1}}$ and $\mathrm{1.31 \cdot 10^{11} cm^{-2}}$ at 1.5K. The surface gates define two quantum dots, with an associated quantum point contact (qpc) gate each. The dot to the left is the thermometer dot and is used to measure the temperature of the 2DEG using the non-invasive method described below. Transport through the right dot is used to verify the temperature obtained from the left dot. The quantum dots were designed to have a radius of 150 nm, and are measured to have a charging energy of approximately 1.5 meV, and first excited states of about 400 $\mathrm{\mu eV}$ above the ground state. 

Measurements were performed in a He$_3$/He$_4$ dilution refrigerator at a base (lattice) temperature of 60 mK. Each quantum dot was set up using three gates. A fourth gate (the qpc gate) was used to define a narrow 1-D channel adjacent to the thermometer dot. The input offset bias of a current to voltage amplifier ($\approx$ 30 $\mathrm{\mu V}$) was used to drive current through this narrow channel. The resulting voltage was recorded on a digital multimeter. A bias of 100 $\mathrm{\mu V}$ was applied across the top dot, and the current through it was measured using an electrometer. The gates colored red in Figure \ref{fig:schematic}(a) were used to vary the energy levels of the two dots through their respective bias windows.

\begin{figure}
\includegraphics{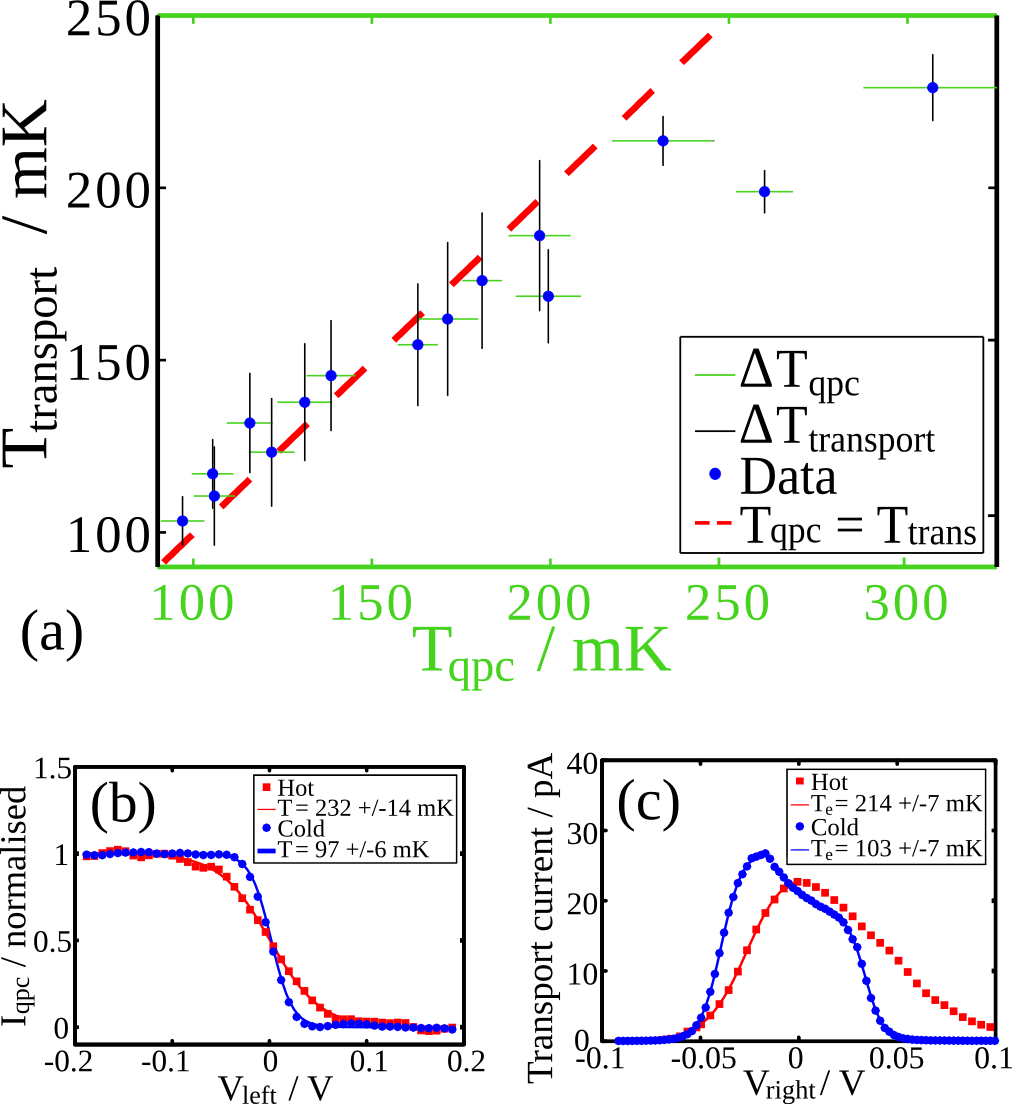}
\caption{Comparing the two thermometry methods as the lattice temperature is varied. (a) Temperatures obtained from transport measurements plotted against those extracted from the qpc signal. The corresponding errors are marked by vertical and horizontal bars respectively. For clarity of representation, the temperature range was divided into equal bins 7.5 mK in size, and data in each bin was averaged.  The dashed red line is the best fit to data (the three warmest points have been excluded because the transport dot is not well defined for higher temperatures). (b),(c) The  qpc signal, and the transport current as a function of the voltages on the respective plunger gates ($\mathrm{V_{left}}$ and $\mathrm{V_{right}}$), for two different temperatures, and the corresponding best fits. The temperatures obtained for the warm (red squares) and cold (blue circles) readings are, (b) from the qpc signal: 232$\pm$14mK, 97$\pm$6mK, and (c) from the transport signal: 214$\pm$7mK, 103$\pm$7mK respectively.}
\label{fig:comparison}
\end{figure}

The left dot is used as a thermometer to measure the temperature of the two-dimensional electron gas (Figure \ref{fig:schematic}(a), inset). The one-dimensional channel (qpc) defined adjacent to it has a conductance quantized in units of $\mathrm{2e^2/h}$, each step corresponding to transport through the next 1-D sub-band \cite{thomas}. The resistance of the channel while on a riser between steps is extremely sensitive to its immediate electrostatic environment (we assume this to be linear for a small change in potential). When an electron enters the dot, the electrostatic environment changes, and a dip is registered in the detector current \cite{charles}. 

The height and width of the right (IN) and left (OUT) barriers of the thermometer dot determine its tunnel-coupling to the source and drain ohmics on either side. Figure \ref{fig:schematic}(b) is a diagram of the current through the qpc (differentiated with respect to the voltage on the plunger gate) as a function of the two barriers. In the top right corner of this figure, the two barriers are equal and are individually so high that no measurable transport current flows through the dot. The tunneling rate at the lowest measurable current is about 600 kHz which corresponds to a negligible lifetime broadening of the states in the dot ($\Gamma \ll $ 1 mK).

In the lower right corner of the map, electrons can tunnel between the dot and the 2DEG on the right at a frequency much less than 600 kHz, and much greater than the measurement sample frequency of 5Hz. As the voltage on the IN gate is made increasingly negative, electrons are emptied out of the dot one by one. As we trace one transition towards the upper left corner, the dot becomes heavily isolated from both the left and right sides, and electrons can leave the the dot only very slowly (on the order of tens of seconds), and we observe a time-dependent scatter of the steps in current when we sweep the gate voltage relatively quickly in one direction (the direction of the scatter depends on the direction of the sweep). In the upper left corner, the dot is open to the 2DEG on the left side, and isolated from the 2DEG on the right. Our temperature measurements were made using the last transition in the lower right corner of the figure (the red square in Figure \ref{fig:schematic}(b) approximately marks the configuration used).

As the energy level of the dot is swept past the thermally broadened chemical potential of the source 2DEG, the occupation probability of electrons in the dot changes. The response of the detector current is directly proportional to this charge occupation probability, which is in turn proportional to the probability of the electron tunneling into the dot from the electron gas. Since the electrons in the two-dimensional electron gas have a constant density of states, their energy spectrum is entirely described by the Fermi-Dirac distribution ($\mathrm{f(\epsilon)=1/(1+e^{-(\epsilon-\mu)/kT})}$, where $\mu$ is the chemical potential of the gas, T its temperature, and k the Boltzmann constant). A scaled Fermi Dirac distribution can thus be fitted to the measured detector current, and the electron temperature can then be extracted from the fit.

In real measurements, a linear background (which models the capacitive effect of the plunger gate voltage on the width of the qpc) was subtracted from the measured current \cite{theory}: 
\begin{equation}
I(V_g)=\frac{n}{1+e^{-\alpha_g(V_g-u_C)/kT}}+lV_g+C
\label{eq:form}
\end{equation}
(where $V_g$ is the voltage on the plunger, $\alpha_g$ is the lever arm of the plunger gate, $l$ is the linear coefficient, $u_C$ is the horizontal offset, and $C$ is the vertical offset which indicates the position of the step along the riser). To reduce the effects of noise, several scans can be averaged, and a Fermi Dirac function fitted to the resulting curve to extract the electron temperature (Figure \ref{fig:schematic}(c)). A single low-noise sweep (which takes 3 minutes) can be used to extract the temperature with errors as low as 3 mK. 

Since this method of measuring the electron temperature has not been used before, the results obtained from it were compared with the temperatures obtained by measuring finite bias Coulomb-blockaded transport in the right dot (Figure \ref{fig:comparison}). The current through a quantum dot in the Coulomb blockade regime is the sum of contributions from the source and drain, each of which is the integral of the product of the appropriate Fermi Dirac distribution, and the transmission resonance of the dot \cite{beenakker}. For low source-drain biases, the resulting linehape is a peak whose width is determined by both the lifetime broadening of the dot's transmission resonance, and the thermal broadening of the source and drain. 

For a high source-drain bias, the difference between the contributions from the source and drain gives rise to a `top-hat' shape, with the width of the top hat proportional to the bias window, and the height of the top hat determined by the tunneling rates to the two reservoirs (Figure \ref{fig:comparison}(c)). In our experiments, the top hat is sloped because of asymmetric tunnel barriers to the source and drain, and the computational model therefore fitted to the right and left sides separately. The width of the sloping sides of the top-hat was determined only by the thermal broadening of the source and drain, and not by the lifetime broadening in the dot, which is estimated to be 6 mK for a transport current of 20pA ($\Gamma \approx hI/e$). We expect the temperatures of the two reservoirs to be equal, and the final value obtained from each scan was a calculated average of the two temperatures extracted from the left and right sides (for warmer temperatures, a leakage current to the right meant that only the left side was used).

In the experiment, the energy levels of the left and right dots were simultaneously swept through their bias windows using the plunger gates $\mathrm{V_{left}}$ and $\mathrm{V_{right}}$ respectively (colored red in Figure \ref{fig:schematic}(a)). A temperature was extracted from each sweep of the qpc and transport signals (Figure \ref{fig:comparison}(b),(c)), and the two values were compared as the lattice temperature of the fridge was decreased from 300mK to the base temperature ($\approx$ 60mK). Both methods were found to agree well over the test temperature range (Figure \ref{fig:comparison}(a)). At warmer temperatures, transport through the right dot consistently gives lower temperatures than those extracted from the qpc signal of the left dot. This coincides with the onset of a leakage current to the source which means that the quantum dot is no longer well defined, and that the fitting procedure described above becomes progressively less accurate. We can conclude that the tuning of the quantum dot used in transport measurements is temperature dependent and needs to be optimized. In contrast, the qpc thermometer is sufficiently isolated from the electron gas that the same tuning can be used for the entire temperature range. Further, because the qpc signal is up to four times greater than the transport signal (Figures \ref{fig:schematic}(c), and \ref{fig:comparison}(c) respectively), the qpc thermometer is much more sensitive than the transport thermometer for a similar experimental set up.

In this work, we have demonstrated a non-invasive scheme for determining the temperature of the 2DEG in a GaAs/AlGaAs heterostructure, which can be run simultaneously with and independently of another operating device. It consists of monitoring the charge occupation of a quantum dot weakly tunnel-coupled to the 2DEG, using a quantum point contact. Electronic temperatures between 97 mK and 307 mK were measured, and agree with those obtained using the orthodox method of extracting temperature from transport measurements through a quantum dot. In our experiment, the upper bound on this technique was limited by telegraphic noise in the wafer (which rendered the qpc signal indistinguishable at higher temperatures), and is approximately 0.5 K. This could be overcome by using a wafer less susceptible to charge fluctuations, and in theory the upper bound is determined only by the sub-band spacing (the width) of the qpc channel. The lower bound on this technique is determined by the lifetime broadening of the single particle states of the thermometer quantum dot, and the time available for a measurement sweep. The smaller lower bound on this non-invasive method resulting from remotely measuring the occupation of a quantum dot means that it can be used to measure lower temperatures than those measurable using Coulomb blockaded transport. The technique is expected to be robust in the presence of a magnetic field, and shows promising potential to become an easy and accurate thermometry standard.

The authors acknowledge support from the EPSRC (Grant EP/K004077/1), and Toshiba Research Europe. AM was supported by the Dr Manmohan Singh Scholarship (2009-12) from St. John's College, Cambridge.

%

\end{document}